\documentclass[aps,prb,twocolumn,showpacs,amsmath]{revtex4}

\usepackage{latexsym}
\usepackage{amssymb}
\usepackage{bm}
\usepackage{graphicx}
\def\stackunder#1#2{\mathrel{\mathop{#2}\limits_{#1}}}
\newcommand{\pdag}{{\phantom{\dagger}}}
\newcommand{\bq}{\begin{equation}}
\newcommand{\eq}{\end{equation}}
\newcommand{\bn}{\begin{eqnarray}}
\newcommand{\en}{\end{eqnarray}}

\begin{document}

\title{Photon-Phonon-assisted tunneling through a single-molecular quantum dot}

\author{Bing Dong$^{1,2}$, H. L. Cui$^{1,3}$, and X. L. Lei$^{2}$} 
\affiliation{$^{1}$Department of Physics and Engineering Physics, Stevens Institute of 
Technology, Hoboken, New Jersey 07030 \\
$^{2}$Department of Physics, Shanghai Jiaotong University,
1954 Huashan Road, Shanghai 200030, China \\
$^{3}$School of Optoelectronics Information Science and Technology, Yantai University, 
Yantai, Shandong, China}

\begin{abstract}

Based on exactly mapping of a many-body electron-phonon interaction problem onto a 
one-body problem, we apply the well-established nonequilibrium Green function technique 
to solve the time-dependent phonon-assisted tunneling at low temperature through a 
single-molecular quantum dot connected to two leads, which is subject to a microwave 
irradiation field. It is found that in the presence of the electron-phonon interaction 
and the microwave irradiation field, the time-average transmission and the nonlinear 
differential conductance display additional peaks due to pure photon absorption or 
emission processes and photon-absorption-assisted phonon emission processes. The 
variation of the time-average current with frequency of the microwave irradiation field 
is also studied.       

\end{abstract}

\maketitle

\section{Introduction}

The role of inelastic scattering in electron transport through mesoscopic devices is an 
active topic of current theoretical and experimental 
interest.\cite{Chen,Park,Ventra,Zhitenev,Fujisawa,Turley,Brandes,Bose,Wingreen,Lundin,Fl
ensberg} Thanks to the modern advanced nanostructure techniques, we can build 
semiconductor quantum dots (QDs) with easy control of its size and shape. Recent 
experiments found that the electron-phonon interaction (EPI) for longitudinal optical 
phonons in semiconductor QD can be substantially enhanced due to the quantum confinement 
effect, and also revealed the importance of the EPI in transport measurements through 
this kind of single electron transistor. On the other hand, a growing interest has been 
payed recently to electrical transport through a very small single molecular device. In 
contrast to the semiconductor QD, the molecular material possess much weaker elastic 
parameters, leading to that their internal vibration degrees of freedom is of 
``low-energy", $1$-$10\,$meV.\cite{Chen,Park,Ventra,Zhitenev} Therefore, when there are 
electrons incident into the molecule through tunneling junction from external 
environment, these low-energy boson modes (phonons) can be easily excited due to the 
strong coupling between these modes and the molecular electronic states, and thus react 
inevitably back to the tunneling of electrons even at very low temperature. For example, 
the interesting experimental results reported in Ref.\,\onlinecite{Park} demonstrated 
that electron transport through a single C$_{60}$ molecule was significantly influenced 
by a single vibrational mode.

This phenomena has provoked a large amount of theoretical work on the problem of 
tunneling through a single level with coupling to phonon modes based on the Fermi golden 
rule,\cite{Turley} the kinetic equation approach,\cite{Brandes,Bose} and the 
nonequilibrium Green's function techniques.\cite{Wingreen,Lundin,Flensberg} Naturally, 
both the phonon energy and the electron-phonon coupling strength are of importance to 
determine transport properties of these single electron transistors. Most theoretical 
attempts to study this transport problem so far have focused on stationary case. As far 
as the energy of boson in molecule is considered to be in the region of GHz to THz, one 
can imagine that more rich physics could be exploited if the device is subject to a 
microwave (MW) irradiation field. It has been reported that the microwave spectroscopy 
is a possible tool to probe the energy spectrum of small quantum systems and to measure 
the decoherence time of the quantum states.\cite{Wiel} So the photon-assisted tunneling 
(PAT) could provide a new way of understanding more essence of the EPI influence on the 
transport properties in molecule.             

The purpose of this paper is to study the time-dependent coherent transport through a 
single-molecular device coupled to a local dispersionless phonon mode by means of 
nonequilibrium Green's function (NGF) under the adiabatic approximation.\cite{Jauho} 
Acutally, when inelastic processes such as phonon emission and absorption are considered 
in the tunneling event, electron scattering becomes a many-body problem involving 
electrons and various excited phonon states of the system. Since the coupling between 
electrons and phonons is strong in the semiconductor and the single-molecuar QD, the 
usual perturbation theory is invalid for dealing with this scattering problem, although 
it has been extensively exploited with enormous success to study the conventional 
transport in bulk material. Recently, a nonperturbative scheme has been proposed by 
Bon\v ca and Trugman\cite{Bonca} for studying the small polaron of the coupled 
electron-phonon system described by the Holstein model\cite{Holstein} and the 
Su-Schrieffer-Heeger (SSH) model.\cite{SSH} The fundamental idea of this method is to 
rewrite the Hamiltonian in terms of the combined electron-phonon Fock space, such can 
exactly map the many-body problem onto an one-body scattering problem. Later on, this 
method has been further applied to the topic of inelastic electron scattering in 
mesoscopic quantum transport in semiconductor QDs and molecular wires under the 
framework of the Landauer-B\"uttiker scattering theory.\cite{Bonca2,Ness} In this paper, 
we will redirect this procedure in terms of the NGF technique and address the inelastic 
PAT in a molecular QD.

The rest of this work is organized as follows. In section II, we introduce the 
Hamiltonian for resonant tunneling, where electrons interact with the boson fields 
localized on the QD or molecule. Then we describe our methodology for studying the 
time-dependent tunneling and computing the time-average transmission and current by 
means of NGF technique. In section III, we give the numerical results and discussions. 
We find some new features in the time-average transmission spectrum associated with the 
combination effects of the photon-phonon emission or absorption. Then we predict that 
this photon-phonon-assisted tunneling can be observe experimentally in the bias 
voltage-dependent differential conductance and in the microwave spectroscopy of dc 
current. Finally, a brief summary is presented in Sec. IV.     

\section{Method and formulation}

\subsection{Model and Hamiltonian}

We consider the system under investigation as the simplest case: a single-site QD 
couples to two noninteracting reservoirs via tunneling and interacts with a 
dispersionless optical phonon localized in this site. It is anticipated that this is 
sufficient to illustrate the main physics for transport through the semiconductor QD and 
molecule in the presence of many boson modes. Moreover, we neglect the spin degree of 
freedom and any effects of electron-electron Coulomb interactions. Therefore, the 
Hamiltonian of this system can be split into three parts, (1) the two isolated leads 
$H_{\rm lead}$, (2) the single-site QD $H_{\rm QD}$, and (3) the tunneling part $H_{\rm 
tl}$:  
\begin{subequations}
\label{Hamiltonian}
\bn
H_{\rm lead}&=& \sum_{\eta, k} \epsilon_{\eta k}(t)c_{\eta k}^{\dagger} c_{\eta 
k}^{\pdag}, \label{lead} \\
H_{\rm QD}&=& \epsilon_{d}(t) c_{d}^{\dagger} c_{d}^{\pdag} + \hbar \omega_{ph} 
b^{\dagger} b^{\pdag} - \lambda c_{d}^{\dagger} c_{d}^{\pdag} (b^{\pdag}+b^{\dagger}), 
\label{QD} \\
H_{\rm tl}&=& \sum_{\eta,k}V_{\eta}(c_{\eta k}^{\dagger} c_{d} + {\rm H.c.}). 
\label{tunneling}
\en    
\end{subequations}
$c_{\eta k}^{\dagger}$ ($c_{\eta k}$), $c_{d}^{\dagger}$ ($c_{d}$) are the creation 
(annihilation) operators for the noninteracting electrons with momentum $k$ in the 
$\eta$ ($=L/R$) lead, and for the electronic state at the single-site, respectively. The 
role of the MW fields irradiated on the whole system can be described by a rigid shift 
of the single-electron energy spectrum under the adiabatic approximation:\cite{Jauho} 
$\epsilon_{\eta k}(t)=\epsilon_{\eta k}^0+v_{\eta}(t)$ ($\eta=L,R$) and 
$\epsilon_{d}(t)=\epsilon_{d}+v_{d}(t)$, $\epsilon_{\eta k}^0$ and $\epsilon_{d}$ are 
the time-independent single electron energy without MW fields, $v_{\eta/d}(t)$ is the 
time-dependent part, $v_{\eta/d}(t)=v_{\eta/d}\cos \Omega t$, with $v_{\eta/d}$ being 
the irradiation strength in different elements of the device and $\Omega$ being the 
frequency of the MW field. $\hbar \omega_{ph}$ is the energy of the dispersionless 
phonon mode and $\lambda$ is the on-site EPI constant. In the present work, we neglect 
the direct coupling between the MW field and the boson field. $V_{\eta}$ stands for the 
tunneling coupling between the dot and $\eta$th lead.      

Obviously, this problem described by the above Hamiltonian (\ref{Hamiltonian}) is a 
many-body problem involving the phonon emission and absorption when the electron tunnels 
through the central region. Following the pioneer work of Bon\v ca and Trugman, we can 
expand the electron states in the single-site onto the polaron eigenstates, the 
direct-product states of the single-electron states and the phonon Fock states,
\bq
|0,n\rangle=c_{d}^{\dagger} \frac{(b^{\dagger})^n}{\sqrt{n!}}|0\rangle, \quad (n\geq 0)
\eq
which means that the electron on the site $0$ is accompanied by $n$ number of the phonon 
$\hbar \omega_{ph}$ ($|0\rangle$ is the vacuum state). After performing this 
representation, the many-body on-site Hamiltonian (\ref{QD}) can be mapped onto an 
one-body one:\cite{Bonca,Bonca2,Ness}
\bn
\widetilde {H}_{\rm QD} &=& \sum_{n\geq 0} [ (\epsilon_{d}+n\hbar \omega_{ph}) 
|0,n\rangle \langle 0,n| \cr
&& \hspace{-1cm} - \lambda \sqrt{n+1}(|0,n+1\rangle \langle 0,n| + |0,n\rangle \langle 
0,n+1|) ].
\en    
Similar expansion of electron states in the two noninteracting leads with the combined 
space $|\eta k,n\rangle=c_{\eta k}^{\dagger} \frac{(b^{\dagger})^n}{\sqrt{n!}}|0\rangle$ 
will change the simple $\eta$ lead Hamiltonian (\ref{lead}) to a pseudo-multi-channel 
model labeled by the phonon quanta $n$ with a weight factor $P_n=(1-e^{-\beta 
\omega_{ph}})e^{-n\beta \omega_{ph}}$, which is the statistic probability of the phonon 
number state $|n\rangle$ at the finite temperature $T$ ($\beta=1/k_{\rm B}T$). Notice 
that the formula $\sum_{n}P_n=1$ guarantees the statistic properties of the leads 
unchanged (see the details in the following subsection).          

Finally, the tunneling part Eq.~(\ref{tunneling}) can be also rewritten in terms of this 
basis set. Considering the presumption presented here that there is no exchange between 
electrons and phonons in the tunneling process, only hopping of the electron states are 
involved in the effective tunneling Hamiltonian:
\bq
\widetilde{H}_{\rm tl}=\sum_{\eta,k,n}V_{\eta n}(|\eta k,n\rangle \langle 0,n|+{\rm 
H.c.}),
\eq
where $|\eta k\rangle$ denotes the electron state with momentum $k$ in the $\eta$ lead. 
It should be noted that this presumption is rational because the high-order tunneling 
processes accompanied by the phonon emission and absorption are much weaker than the 
direct tunneling events. And in fact, inclusion of these high-order processes in the 
calculations is an instant task under the present theoretical framework. $V_{\eta n}$ is 
the coupling between the $n$th pseudo-channel in the $\eta$ lead and the QD.     

\begin{figure}[htb]
\includegraphics [width=5cm,height=4cm,angle=0,clip=on] {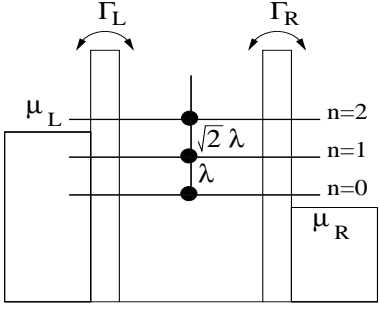}
\caption{Schematic description of the inelastic scattering problem for a single site 
with on-site EPI. Each phonon state of the site along with the Bloch state of the 
electron in the lead can be visualized as a pseudo-channel labeled by $n$, which is 
connected to the two leads with the hopping parameters $\Gamma_{L/R}$ in the wide band 
limit. These channels are connected vertically by the EPI $\lambda$.} \label{fig1}
\end{figure}

We depict the many-body tunneling process after remodeling EPI system in terms of the 
combined electron-phonon Fock space: An electron incident in the pseudo-channel labeled 
by a particular phonon number $n$ in the left lead can transport elastically to the 
central region, then excites or absorbs $m$ numbers of phonons and exits inelastically 
into the $(n\pm m)$th channels in the left and right leads; or experiences no exchange 
with phonon and leaves elastically at the same channel in both leads. Fig.\,1 shows a 
graphical illustration for this description.

So far, all theoretical solutions in the literature for the phonon-assisted tunneling 
under above mapping Hamiltonian are based on the Landauer-B\"uttiker scattering 
formulation.\cite{Bonca,Bonca2,Ness} In this paper, in order to utilize the 
well-established NGF technique to deal with the time-dependent phonon-assisted 
tunneling, we first define these Dirac brackets as operators:
\bq
c_{d n}^{\dagger}=|0,n\rangle, \quad c_{\eta k n}^{\dagger}=|\eta k,n\rangle,
\eq
then rewrite the mapping Hamiltonian in the operator representation:
\bn
H_{\rm eff}&=&\sum_{\eta, k, n} \epsilon_{\eta k n}(t)c_{\eta k n}^{\dagger} c_{\eta k 
n}^{\pdag}+ \sum_{n} [\epsilon_{dn}(t) c_{d n}^{\dagger} c_{d n}^{\pdag} \cr
&&- \lambda \sqrt{n+1}(c_{d n+1}^{\dagger} c_{d n}^{\pdag} + {\rm H.c.}) ] \cr
&&+ \sum_{\eta,k,n}V_{\eta n}(c_{\eta k n}^{\dagger} c_{d n}^{\pdag}+{\rm H.c.}), 
\label{Hamiltonianeff}
\en       
with $\epsilon_{\eta k n}=\epsilon_{\eta k}(t)+n\hbar \omega_{ph}$ and 
$\epsilon_{dn}=\epsilon_{d}(t)+n\hbar \omega_{ph}$ ($n\geq 0$). Note that these 
pseudo-Fermi operators in the effective Hamiltonian are bilinear. Finally, the explicit 
anticommunicators of these operators can be easily determined:
\bq
\{ c_{dn}^{\pdag}, \, c_{dm}^{\dagger}\}=\delta_{nm}, \quad \{ c_{\eta k n}, \, c_{\eta' 
k' m}^{\dagger}\}=\delta_{\eta \eta'}\delta_{k k'}\delta_{nm}. \label{comm}
\eq   

\subsection{Mathematical Formulation} 

In the subsection, we derive the general formula of the time-dependent transmission 
$T_{\rm tot}(t)$ and the time-average transmission $T_{\rm tot}$ and current $I$ through 
the QD coupled to dispersionless Einstein phonons with frequency $\omega_{ph}$ by using 
the NGF technique.\cite{Jauho} The time-dependent current $J_{Ln}(t)$ from the $n$th 
channel of the left lead to the QD can be calculated from the time evolution of the 
occupation number operator of the electrons in this channel, $N_{Ln}=\sum_{k} 
c_{Lkn}^{\dagger} c_{Lkn}^{\pdag}$, and one readily finds
\bn
&&\hspace{-0.5cm} J_{Ln}(t)=-e\langle \dot N_{Ln} \rangle=-\frac{ie}{\hbar}\langle 
[H_{\rm eff}(t), N_{Ln}]\rangle \cr
&&=\frac{ie}{\hbar}\sum_{k} [V_{Ln} G_{n,Lkn}^{<}(t,t) - V_{Ln}^* \langle 
G_{Lkn,n}^{<}(t,t) \rangle].
\en
Here we define these hybrid retarded (advance) and lesser (greater) GFs $G_{n, \eta 
km}^{r(a),<(>)}(t,t')\equiv \langle \langle c_{dn}(t)| c_{\eta km}^{\dagger}(t') 
\rangle\rangle^{r(a),<(>)}$ as follwos: $G_{n, \eta km}^{r(a)}(t,t')\equiv \pm i \theta 
(\pm t\mp t')\langle \{c_{dn}(t), c_{\eta km}^{\dagger}(t') \}\rangle$, $G_{n,\eta 
km}^{<}(t,t')\equiv i\langle c_{\eta km}^{\dagger}(t') c_{dn}^{\pdag}(t)\rangle$ and 
$G_{n,\eta km}^{>}(t,t') \equiv -i \langle c_{dn}^{\pdag}(t) c_{\eta km}^{\dagger}(t') 
\rangle$; and similar definitions for $G_{\eta km,n}^{r(a),<(>)}(t,t')\equiv \langle 
\langle c_{\eta km}(t)| c_{dn}^{\dagger}(t') \rangle\rangle^{r(a),<(>)}$. With the help 
of the Langreth analytic continuation rules,\cite{Langreth} these hybird GFs can be 
related to the GFs of the QD $G_{mn}^{r(a),<(>)}(t,t')\equiv \langle \langle c_{dm}(t)| 
c_{dn}^{\dagger}(t') \rangle\rangle^{r(a),<(>)}$, where $G_{mn}^{r(a)}(t,t')\equiv \pm i 
\theta (\pm t\mp t') \langle \{c_{dm}(t), c_{dn}^{\dagger}(t') \}\rangle$, 
$G_{mn}^{<}(t,t')\equiv i \langle c_{dn}^{\dagger}(t') c_{dm}^{\pdag}(t)\rangle$, and 
$G_{mn}^{>}(t,t')\equiv -i \langle c_{dm}^{\pdag}(t) c_{dn}^{\dagger}(t') \rangle$. Then 
the time-dependent current $J_{Ln}(t)$ becomes
\begin{widetext}
\bq
J_{Ln}(t)= \frac{e}{\hbar}|V_{Ln}|^2 \sum_{k} \int_{-\infty}^{t} dt_1 \left \{ 
G_{nn}^{<}(t,t_1) g_{Lkn}^a(t_1,t) + G_{nn}^{r}(t,t_1) g_{Lkn}^{<}(t_1,t) 
-g_{Lkn}^{<}(t,t_1) G_{nn}^{a}(t_1,t) - g_{Lkn}^r(t,t_1) G_{nn}^{<}(t_1,t) \right \}, 
\label{tdiln} 
\eq   
\end{widetext}
in which $g_{\eta kn}^{r(a),<(>)}(t,t')$ are the free-electron GFs in the $n$th 
pseudo-channel of the $\eta$ lead without coupling to the central region.\cite{Jauho} 
Employing the property $G^r-G^a\equiv G^{>}-G^{<}$ and Noticing the definitions of these 
GFs, Eq.\,(\ref{tdiln}) can be simplified as
\bn
J_{Ln}(t)&=& \frac{e}{\hbar}|V_{Ln}|^2 \sum_{k} \int_{-\infty}^{t} dt_1 \left \{ 
G_{nn}^{>}(t,t_1) g_{Lkn}^<(t_1,t) \right. \cr
&&\left. - g_{Lkn}^{>}(t,t_1) G_{nn}^{<}(t_1,t) \right \}. 
\label{iLnt} 
\en
In the wide band limit, where the hopping matrix element $V_{\eta n}=V_{\eta}$ is 
independent on energy, one has
\bn
\sum_{k}|V_{\eta n}|^2 g_{\eta kn}^{<}(t,t')&=& i\Gamma_{\eta} \int \frac{d\omega}{2\pi} 
f_{\eta}^n(\omega) F(t,t'), \label{gflesslead} \\
\sum_{k}|V_{\eta n}|^2 g_{\eta kn}^{>}(t,t')&=& -i\Gamma_{\eta} \int 
\frac{d\omega}{2\pi} [1-f_{\eta}^n(\omega)] F(t,t'), \nonumber
\en
with $F(t,t')=e^{-i \omega (t-t')-i \int_{t'}^t d\tau v_{\eta}(\tau)}$, 
$\Gamma_{\eta}=2\pi \sum_{k} |V_{\eta}|^2 \delta(\omega- \epsilon_{\eta k})$ being the 
generalized linewidth function, and $f_{\eta}^n=\{ 1+ e^{(\omega+n\hbar 
\omega_{ph}-\mu_{\eta})/k_{\rm B}T}\}^{-1}$ being the Fermi distribution function of the 
$n$th pseudo-channel in the $\eta$ lead.   

Therefore to compute the time-dependent current, we have to evaluate the Keldysh GFs 
$G_{nn}^{<(>)}(t,t')$. In the following, we first calculate the retarded GF 
$G_{mn}^{r}(t,t')$ in the equation of motion approach. Introducing the gauge 
transformation:\cite{Ng}
\begin{subequations}
\bn
G_{mn}^{r}(t,t')&=& e^{-i\int_{t'}^t d\tau v_{d}(\tau)} \bar 
{G}_{mn}^r(t,t'),\label{gaugfr} \\
G_{\eta km,n}^{r}(t,t')&=& e^{-i\int_{t'}^t d\tau v_{\eta}(\tau)} \bar {G}_{\eta 
km,n}^r(t,t'),
\en
\end{subequations}
one easily finds
\bn
&&\left ( i\frac{\partial}{\partial t}- \epsilon_{d}-m\hbar \omega_{ph} \right ) \bar 
{G}_{mn}^{r}(t,t')=\delta_{mn}\delta(t-t') \cr
&& - \lambda \sqrt{m+1} \bar{G}_{(m+1)n}^{r}(t,t') - \lambda \sqrt{m} 
\bar{G}_{(m-1)n}^{r}(t,t') \cr
&& + \sum_{\eta,k} V_{\eta m} \bar{G}_{\eta k m,n}^{r}(t,t') e^{i\int_{t'}^t d\tau 
\Delta_{\eta}(\tau)},
\en
and
\bn
\left ( i\frac{\partial}{\partial t}- \epsilon_{\eta k}^0-m\hbar \omega_{ph} \right ) 
\bar {G}_{mn}^{r}(t,t') &=& V_{\eta m}^* \bar{G}_{mn}^{r}(t,t') \cr
&& \hspace{-1cm} \times e^{-i\int_{t'}^t d\tau \Delta_{\eta}(\tau)},
\en
with $\Delta_{\eta}(t)=v_{d}(t)-v_{\eta}(t)$. Then we transform the time coordinates to 
energy coordinates following the usual prescription,
\bq
Y(\omega,\omega')=\int dt \int dt' \, Y(t,t')e^{i\omega t-i\omega' t'},
\eq 
and obtain
\bn
&&\hspace{-0.5cm}[\omega-\epsilon_{d}-m\hbar \omega_{ph}-\bar \Sigma_{m}^{r}(\omega)] 
\bar{G}_{mn}^{r}(\omega, \omega') =\delta_{mn}\delta(\omega-\omega') \cr
&& \hspace{-0.5cm} - \lambda \sqrt{m} \bar{G}_{(m-1)n}^{r}(\omega,\omega') - \lambda 
\sqrt{m+1} \bar{G}_{(m+1)n}^{r}(\omega,\omega'),\label{gfrome}
\en
where the retarded self-energy $\bar \Sigma_{m}^{r}(\omega)$ due to the coupling to the 
leads through the same gauge transformation as for the GFs is defined as
\bn
\bar \Sigma_{m}^{r}(\omega)&=& \sum_{\eta,k} |V_{\eta m}|^2 \bar{g}_{\eta 
km}^{r}(\omega,\omega') \delta(\omega-\omega') \cr  
&=&\sum_{\eta,k}\frac{|V_{\eta m}|^2}{\omega-\epsilon_{\eta k m}^0-m\hbar \omega_{ph}+i 
0^{+}}.
\en
In the wide band approximation, we have $\bar \Sigma_{m}^{r} = - \frac{i}{2} (\Gamma_{L} 
+ \Gamma_{R})=-\frac{i}{2}\Gamma$, which leads to $\bar{\Sigma}_{m}^{r}(t,t') 
\longrightarrow -i\frac{i}{2}\delta(t-t')$. Here to avoid unnecessary complications, we 
ignore the EPI induced band narrowing, which will lead to some trivial quantitative but 
no significant qualitative changes for tunneling current.\cite{Lundin} Therefore, it is 
helpful to rewrite the resulting Eq.\,(\ref{gfrome}) in a compact matrix form:
\bq
[(\omega+i\Gamma){\bm I}-{\bm B}] \bar{{\bm G}}^{r}(\omega)={\bm I}, 
\eq
in which ${\bm B}$ is a $N\times N$ ($N=\infty$ denotes the total phonon number) 
symmetrical tridiagonal matrix: $B_{nn}=\epsilon_{d}+(n-1)\hbar \omega_{ph}$, 
$B_{n(n-1)}=-\lambda \sqrt{n-1}$, and $B_{n(n+1)}=-\lambda \sqrt{n}$, ${\bm I}$ is the 
$N$ dimensional unit matrix. A similar result is obtained for the advanced GF  
$\bar{\bm G}^a(\omega)$, and simply we have $\bar{\bm G}^a=[\bar{\bm G}^r]^{\dagger}$ in 
the Fourier space. After performing the reverse Fourier transformation of $\bar {\bm 
G}^{r}(\omega)$, then substituting the resulting $\bar{\bm G}^{r}(t,t')$ into 
Eq.\,(\ref{gaugfr}), one obtains
\bq
{\bm G}^{r}(t,t')=\int \frac{d\omega}{2\pi} \frac{1}{(\omega+i\Gamma){\bm I}-{\bm B}} 
e^{-i\omega (t-t')-i\int_{t'}^t d\tau v_{d}(\tau)}. \label{gfr}
\eq

Now we proceed to calculate the correlation GFs ${\bm G}^{<(>)}(t,t')$. Using the formal 
Keldysh GF technique, they are related to the retarded and advanced GFs as
\bq
{\bm G}^{<(>)}(t,t')=\int dt_1 \int dt_2 {\bm G}^{r}(t,t_1) {\bm \Sigma}^{<(>)}(t_1,t_2) 
{\bm G}^{a}(t_2,t'),
\eq  
with the ``scattering in (out)" Keldysh self-energy ${\bm \Sigma}^{<(>)}$ in the matrix 
form with respect to the phonon number. Because the strongly correlated Hamiltonian 
(\ref{Hamiltonian}) is transformed exactly to an one-body problem 
(\ref{Hamiltonianeff}), these Keldysh self-energies are produced only by tunneling 
coupling to the leads and they are easily derived, in the wide band limit, as
\begin{subequations}
\label{sefen}
\bn
\Sigma_{mn}^{<}(t,t') &=& \sum_{\eta} \Sigma_{\eta m}^{<}(\omega) \, \delta (t-t') 
\delta_{mn} \cr
&=& i \sum_{\eta} \Gamma_{\eta} f_{\eta}^m(\omega) \, \delta (t-t') \delta_{mn}, 
\label{sefen2} \\ 
\Sigma_{mn}^{>}(t,t') &=& \sum_{\eta} \Sigma_{\eta m}^{>}(\omega) \, \delta (t-t') 
\delta_{mn} \cr
&\hspace{-1.5cm}=&\hspace{-1cm}\,\, -i \sum_{\eta} \Gamma_{\eta} [1 - 
f_{\eta}^m(\omega)] \, \delta (t-t') \delta_{mn}. 
\en 
\end{subequations}
With Eqs.\,(\ref{gfr}) and (\ref{sefen}), the Keldysh GF $G_{mm}^{<(>)}(t,t')$ can be 
determined
\bn
G_{mm}^{<(>)}(t,t') &=& \sum_{\eta,n} \int \frac{d\omega}{2\pi} 
e^{-i\omega(t-t')-i\int_{t'}^t d\tau v_{\eta}(\tau)} \cr
&& \hspace{-1cm} \times \Sigma_{\eta n}^{<(>)}(\omega) A_{\eta mn}^{r}(\omega,t) 
[A_{\eta mn}^{r}(\omega,t')]^{*}, \label{gfless}
\en
in terms of the the auxiliary function $A_{\eta mn}^{r}(\omega,t)$:
\bn
A_{\eta mn}^{r}(\omega,t) &=& \int_{-\infty}^{t} dt_1 \int \frac{d\omega'}{2\pi} e^{i 
(\omega'-\omega) (t-t_1)-i \int_{t_1}^t d\tau \Delta_{\eta}(\tau)} \cr
&& \hspace{2.3cm} \times \bar {G}_{mn}^{r}(\omega'). \label{A}
\en

Finally we substitute Eqs.\,(\ref{gflesslead}), (\ref{gfless}) and (\ref{A}) into 
Eq.\,(\ref{iLnt}) to calculate the current $J_{Ln}(t)$. As described in the above 
subsection, the current in the $n$th pseudo-channel of the left lead can be divided into 
two parts: $J_{Ln}^{(-)}(t)$ denotes the component outgoing from the $Ln$ channel to all 
pseudo-channels in both leads while $J_{Ln}^{(+)}(t)$ means the part incoming from all 
pseudo-channels in both leads into this counted channel. It is obvious that the two 
terms bracketed in a pair of parentheses in Eq.\,(\ref{iLnt}) are exactly corresponding 
to the two different tunneling processes, and we will write them down separately in the 
following. After some algebra, we obtain
\begin{subequations}
\bn
J_{Ln}^{(-)}(t)&=& \frac{ie}{h}\Gamma_{L} \sum_{\eta, l} \int d\omega f_{L}^n(\omega) 
\Sigma_{\eta l}^{>}(\omega) |A_{\eta nl}^{r}(\omega,t)|^2, \\
J_{Ln}^{(+)}(t)&=& \frac{ie}{h}\Gamma_{L} \sum_{\eta, l} \int d\omega 
[1-f_{L}^n(\omega)] \Sigma_{\eta l}^{<}(\omega) |A_{\eta nl}^{r}(\omega,t)|^2. \cr
&&
\en
\end{subequations}
Another important point we must emphasize is that an appropriate weight factor should be 
added for any components: product by the probability $P$ belonging to the pseudo-channel 
electrons inject from. Therefore, the total time-dependent current flowing through the 
left lead is as a sum over all pseudo-channels in the left leads:
\bn
J(t)&=& \frac{e}{h} \Gamma_{L} \int d\omega \sum_{\eta,n,l} |A_{\eta 
nl}^{r}(\omega,t)|^2 \hspace{2cm}\cr
&& \hspace{-2cm}\times \left \{ P_n f_{L}^n(\omega) \Gamma_{\eta} 
[1-f_{\eta}^l(\omega)]- P_l \Gamma_{\eta} f_{\eta}^l(\omega) [1-f_{L}^n(\omega)] \right 
\}.
\en
Taking into account the fact that ${\bm A}_{\eta}^{r}(\omega,t)$ is a symmetric matrix, 
the total current $J(t)$ is simplified as
\begin{subequations}
\bn
J(t)&=&\frac{e}{h}\Gamma_{L}\Gamma_{R} \int d\omega \sum_{n,l} |A_{R nl}^{r} 
(\omega,t)|^2 \hspace{2cm}\cr
&& \hspace{-1.5cm}\times \left \{ P_n f_{L}^n(\omega) [1-f_{R}^l(\omega)] - P_l 
f_{R}^l(\omega) [1-f_{L}^n(\omega)] \right \}.
\en   
We can also compute only the elastic contribution to the total current by imposing the 
constraint of elastic tunneling $n=l$:
\bq
J_{\rm el}(t)=\frac{e}{h}\Gamma_{L}\Gamma_{R} \int d\omega \sum_{n} |A_{R 
nn}^{r}(\omega,t)|^2 P_n [f_{L}^n(\omega) - f_{R}^n(\omega)].
\eq
\end{subequations}
Similarly, we can define the time-dependent total transmission $T_{\rm tot}(\omega,t)$ 
and the elastic transmission $T_{\rm el}(\omega,t)$ as:
\begin{subequations}
\bn
T_{\rm tot}(\omega,t)&=& \Gamma_{L} \Gamma_{R} \sum_{n,l} P_n |A_{R 
nl}^{r}(\omega,t)|^2,\\
T_{\rm el}(\omega,t)&=& \Gamma_{L} \Gamma_{R} \sum_{n} P_n |A_{R nn}^{r}(\omega,t)|^2.
\en
\end{subequations}

In order to check our model and theory, we inspect the derived current and transmission 
probability formulae in two special cases: 1) without the MW irradiation field; 2) 
absence of the EPI. First if all time-varying parts in energy are zero $v_{L/R,d}=0$, 
one has $A_{\eta nl}^{r}(\omega,t)=\bar{G}_{nl}^{r}(\omega)$. Then 
$t_{nl}(\omega)=\Gamma_{L} \Gamma_{R} |\bar{G}_{nl}^{r}(\omega)|^2$ describes the 
transmission probability from $n$th channel to the $l$th channel according to the 
Lee-Fisher formulation.\cite{Lee} The total transmission becomes $T_{\rm 
tot}(\omega)=\sum_{n,l} P(n) t_{nl}(\omega)$, and the time-independent total current is
\bn
J&=&\frac{e}{h}\Gamma_{L}\Gamma_{R} \int d\omega \sum_{n,l} |\bar{G}_{nl}^{r}(\omega)|^2 
\hspace{2cm}\cr
&& \hspace{-1.5cm}\times \left \{ P_n f_{L}^n(\omega) [1-f_{R}^l(\omega)] - P_l 
f_{R}^l(\omega) [1-f_{L}^n(\omega)] \right \}.
\en   
Both of them are in consistent with the previous results based on the 
Landauer-B\"uttiker formulation. Further, if there is no EPI, $\lambda=0$, the matrix 
${\bm B}$ reduces to a diagonal matrix with the nonzero element 
$B_{nn}=\epsilon_{d}+(n-1)\hbar \omega_{ph}$. Easily, we obtain
\bn
J=J_{\rm el}&=&\frac{e}{h}\Gamma_{L}\Gamma_{R} \int d\omega \sum_{n}  
\frac{1}{\mid\omega-\epsilon_{d}-n\hbar \omega_{ph}+i\Gamma \mid^2}\cr
&& \times P_n [f_{L}^n(\omega)- f_{R}^n(\omega)].
\en   
Recalling $\sum_{n} P_{n}=1$, simple algebra calculation gives
\bq
J=\frac{e}{h}\Gamma_{L}\Gamma_{R} \int d\omega  \frac{1}{\mid\omega-\epsilon_{d}+i\Gamma 
\mid^2} [f_{L}(\omega)- f_{R}(\omega)].
\eq   
This is the famous time-independent current formula for the non-interacting 
resonant-level model.  

We return to the photon-phonon assisted current calculation. Because the easily measured 
current is the dc component, we must compute the time-average current $I=\langle 
J(t)\rangle$. Define the time-average of a time-dependent object $Y(t)$ as
\bq
\langle Y(t) \rangle =\stackunder{t_{p}\rightarrow \infty}{\rm Lim} \frac{1}{t_{p}} 
\int_{-t_{p}/2}^{t_{p}/2} dt\, Y(t).
\eq
Of course, if $Y(t)$ is a periodic function of time, it is sufficient to average over 
the period. For the harmonic time modulation interested here, the integration over $t_1$ 
in the auxiliary function $A_{\eta mn}^{r}(\omega,t)$ is easily to carried out:
\bn
A_{\eta mn}^{r}(\omega,t)&=& \exp \left [ -i \frac{\Delta_{\eta}}{\Omega} \sin \Omega t 
\right ] \cr
&&\hspace{-0.5cm}\times \sum_{l=-\infty}^{\infty} J_{l}(\frac{\Delta_{\eta}}{\Omega}) 
e^{il\Omega t} \bar{G}_{mn}^{r}(\omega-l \Omega),
\en
where $J_{l}(x)$ is the $l$th order Bessel function of the first kind and 
$\Delta_{\eta}=v_{d}-v_{\eta}$ is the difference of the intensities of the MW field on 
the QD and the $\eta$ lead. Therefore, the time averaging of the current and the 
transmission $T_{\rm tot}(\omega)=\langle T_{\rm tot}(\omega,t) \rangle$ will be
\bn
I &=& \frac{e}{h}\Gamma_{L}\Gamma_{R} \int d\omega \sum_{n,m} \sum_{l=-\infty}^{\infty} 
J_{l}^2(\frac{\Delta_{R}}{\Omega}) |\bar{G}_{nm}^{r}(\omega-l\Omega)|^2 \hspace{0.5cm} 
\cr
&&\hspace{-0.8cm} \times \left \{ P_n f_{L}^n(\omega) [1-f_{R}^m(\omega)] - P_m 
f_{R}^m(\omega) [1-f_{L}^n(\omega)] \right \}, \label{itot}
\en
and
\bq
T_{\rm tot}(\omega)= \Gamma_{L} \Gamma_{R} \sum_{n,m} P_n \sum_{l=-\infty}^{\infty} 
J_{l}^2(\frac{\Delta_{R}}{\Omega}) |\bar{G}_{nm}^{r}(\omega-l\Omega)|^2, \label{ttot}
\eq
and two similar expressions for $I_{\rm el}=\langle J_{\rm el}(t)\rangle$ and $T_{\rm 
el}(\omega)=\langle T_{\rm el}(\omega,t)\rangle$ if setting $n=m$.  
 
\section{Numerical results and discussions}

In this section, we start to numerically study the properties of the time-average 
transmission and current through a single-molecular QD under MW irradiation fields, on 
the basis of Eqs.\,(\ref{itot}) and (\ref{ttot}). For simplicity, in our calculation we 
assume the tunneling coupling between the molecular QD and the two leads to be symmetric 
$\Gamma_{L}=\Gamma_{R}=\Gamma$, and the applied MW fields to be also symmetric 
$v_{L}=v_{R}=v$ and then $\Delta_{L}=\Delta_{R}=\Delta$. We set the phonon energy as the 
unit of energy throughout the rest of the paper and choose the Fermi levels of the two 
leads as the reference of energy $\mu_{L}=\mu_{R}=0$ at equilibrium condition.

A considerably wide range for the values of the phonon energy has been estimated for 
various mesoscopic systems in some experimental papers, from $10\,\mu$eV to $10\,$meV, 
in addition to a very weak coupling to the leads $\Gamma\sim 
1\,\mu$eV.\cite{Chen,Park,Ventra,Zhitenev,Fujisawa} In the literature, some different 
values for the phonon energy $\omega_{ph}$, EPI constant $\lambda$, and the tunneling 
constant $\Gamma$ have been used for theoretical 
analysis.\cite{Brandes,Bose,Wingreen,Lundin,Flensberg,Bonca2,Ness} In the present paper, 
we will particularly choose these parameters as typical values for the purpose of 
providing a satisfactory interpretation for the obtainable experimental results about 
low-temperature tunneling in the single-molecular QD. We choose $\lambda=0.5$, 
$\Gamma=0.04$, and $\omega_{ph}=1$ as the unit of energy. In the following calculations, 
the temperature is assumed to be $T=0.04$.

Worth pointing out that the present approach is an exact method to deal with the 
electron-phonon coupled system with arbitrary coupling strength if the maximum number of 
phonons $N_{ph}$ involved in calculation is counted up to infinite. Unfortunately it is 
impossible in real calculation and we have to truncate this number to a finite value as 
long as the computation convergence is reached with any desired accuracy. The 
appropriate value of $N_{ph}$ depends on the energy of the Einstein phonon mode, the EPI 
constant, and the temperature of the system under investigation. Concerned with these 
parameters in the present paper, we choose $N_{ph}=5$ to obtain results with higher than 
$1\%$ accuracy.

\subsection{Time-average transmission}     

First we examine the time-average transmission. In Fig.\,2 we plot the total and elastic 
time-average transmissions as a function of the incoming electron energy $\omega$ for 
the case of QD bare level $\epsilon_d=0$ under different MW irradiation fields: the 
frequency $\Omega$ of the MW field is (b) lower than, (c) equal to, and (d) higher than 
the frequency of the dispersionless phonon $\omega_{ph}$ for the case of a weak MW 
intensity $\Delta/\Omega<1$ ($\Delta=0.2$). For comparison, we also plot the results 
without irradiation in Fig.\,2(a). The pair of numbers $(n,m)$ located near the peaks 
means numbers of the phonon and photon electrons exchange with the heat bath and the MW 
field, respectively, when electrons tunnel through the central region. The positive 
(negative) of the number denotes emission (absorption) of quanta. In absence of 
irradiation fields, three important features can be observed with respect to the EPI: 1) 
The central peak of the transmission shifts from the position $\omega=\epsilon_{d}$ to 
$\epsilon_{d}-\lambda^2/\omega_{ph}$ with a slightly suppressed amplitude (smaller than 
$1$) due to the polaron effect; 2) The positions of the side peaks corresponding to 
emission of $n$ phonons are approximately given by 
$\omega=\epsilon_{d}-\lambda^2/\omega_{ph}+n\hbar \omega_{ph}$, meaning that the new 
pseudo-channel is opened and participates contribution to tunneling; 3) There are no 
peaks in the left side of the central peak because before tunneling there is no 
available phonon for absorption at low temperature. These observations are just in 
agreement with the previous theoretical results.\cite{Wingreen,Lundin,Bonca2,Ness}                          

\begin{figure}[htb]
\includegraphics [width=8cm,height=6.cm,angle=0,clip=on] {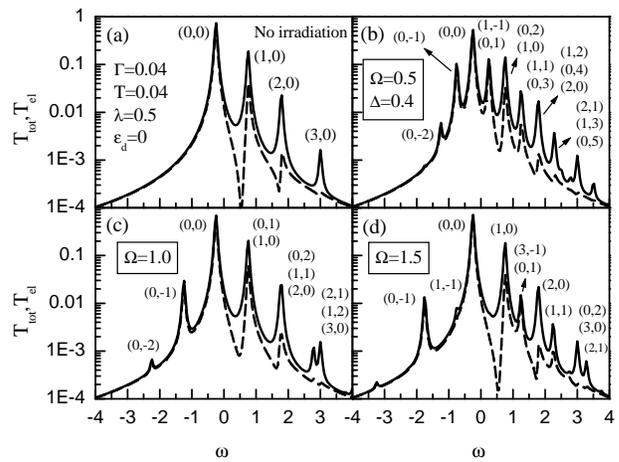}
\caption{Time-average transmission probability as a function of the incident electron 
energy under different irradiations: (a) No irradiation; (b) The frequency of the MW 
field $\Omega=0.5$; (c) $\Omega=1$; (d) $\Omega=1.5$. The intensity of the irradiation 
is $\Delta=0.4$. The solid line is the total transmission $T_{\rm tot}$, and the dashed 
line is the elastic part $T_{\rm el}$. The pair of numbers $(n,m)$ near each peak 
denotes the numbers for emission (positive number) or absorption (negative number) of 
phonon and photon, respectively, involved in the tunneling process. The other parameters 
used in calculation are: $\Gamma=0.04$, $\epsilon_{d}=0$, $\omega_{ph}=1.0$, and 
$\lambda=0.5$. The temperature is set to $T=0.04$.} \label{fig2}
\end{figure}

In the presence of irradiation MW field, it is clear to observe that more rich peaks 
have been observed in the transmission when the involved quanta pair $(n,m)$ satisfies 
$\omega=\epsilon_{d}-\lambda^2/\omega_{ph}+n\hbar \omega_{ph}+m \hbar \Omega$ for the 
given frequency of MW field. In contrast to the phonon, pure absorption peaks of photon 
can be seen in negative incident energy region due to the fact that photon is 
indepdendent of temperature. For the case of lower MW frequency $\Omega=0.5<\omega_{ph}$ 
[Fig.\,2(b)], there appears a new peak in the middle of the zero-phonon-peak $(0,0)$ and 
the first satellite-phonon-peak $(1,0)$, which is attributed to the following two 
processes: $(0,1)$ pure emission of one-photon and $(1,-1)$ emission of phonon assisted 
by absorption of one-photon. And these phonon peaks, such as $(1,0)$ and $(2,0)$, could 
be also accompanied by emission of photon and photon-assisted multi-phonon processes, 
leading to obvious enhancement of the contribution of the elastic channel to 
transmission, as shown in Fig.\,2(b) and (c) for the cases of irradiation fields with 
low frequencies. If the frequency of the MW field is higher than the frequency of 
phonon, no photon-assisted processes take place between the main phonon peak $(0,0)$ and 
the first phonon peak $(1,0)$ because there is insufficient energy to excite the 
high-frequency photon. As a result, the contribution of the elastic channel keeps low. 
More interestingly, we observe a weak emission of one-phonon at the negative energy 
region with the help of absorption of one-photon even under this small intensity of 
irradiation.     

\begin{figure}[htb]
\includegraphics [width=8.5cm,height=7.cm,angle=0,clip=on] {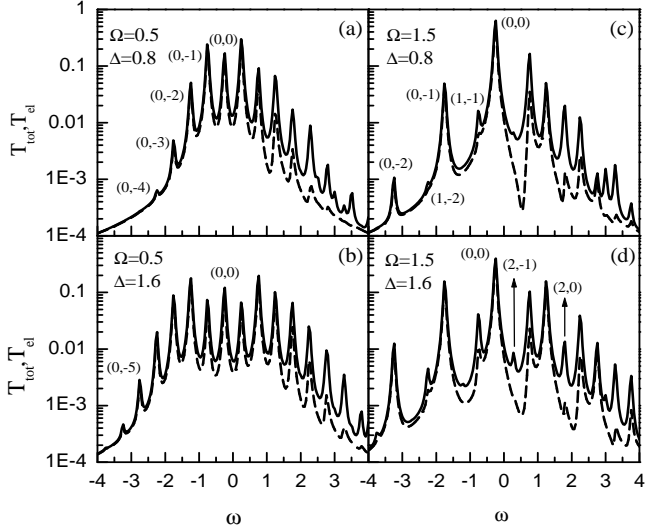}
\caption{Time-average transmission probability as a function of the incident electron 
energy under irradiations with frequencies lower [(a, b): $\Omega=0.5$] and higher [(c, 
d): $\Omega=1.5$] than the frequency of the Einstein phonon. The intensities of the 
irradiations are $\Delta=0.8$ (a, c) and $1.6$ (b, d). The other parameters are the same 
as in Fig.\,2.} \label{fig3}
\end{figure}

Increasing intensity of the MW field will enhance the photon-assited processes. In 
Fig.\,3 we plot the time-average transmissions under the MW fields of frequencies 
$\Omega=0.5$ and $1.5$ with stronger intensities $\Delta=0.8$ and $1.6$. Obviously, we 
observe more peaks with more high amplitudes involving emission and absorption of 
photons, as well as suppression of the pure phonon-assisted processes. For the strongest 
intensity $\Delta=1.6$ calculated here, the high-frequency photon ($\Omega=1.5$) even 
excites an emission peak involving two-phonon process between the main-phonon and the 
one-phonon peaks, as depicted in Fig.\,3(d).

\subsection{Time-average current and nonlinear differential conductance}     

\begin{figure}[htb]
\includegraphics [width=8.cm,height=10.5cm,angle=0,clip=on] {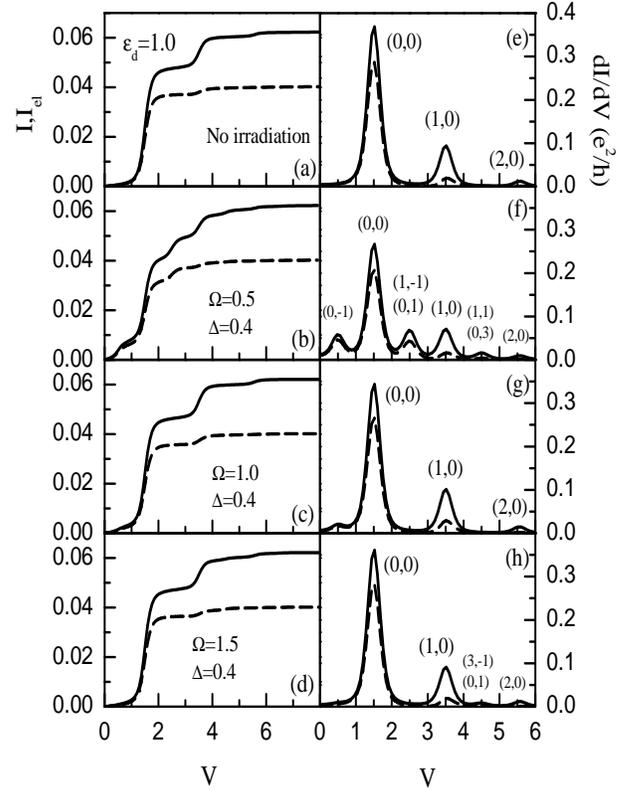}
\caption{The time-average total current $I$ (solid line), elastic current $I_{\rm el}$ 
(dashed line) [(a-d)], and the corresponding differential conductance $dI/dV$ [(e-h)] as 
a function of the applied bias voltage for a single-molecular QD with bare level 
$\epsilon_{d}=1.0$ under different irradiations: (a,e) No irradiation; (b,f) The 
frequency of the MW field $\Omega=0.5$; (c,g) $\Omega=1$; (d,h) $\Omega=1.5$. The 
intensity of the irradiation is $\Delta=0.4$. The other parameters are the same as in 
Fig.\,2.} \label{fig4}
\end{figure}

According to the current formula Eq.\,(\ref{itot}), nonlinear differential 
conductance, defining as the derivative of the time-average current with respect to 
the bias voltage $dI/dV$ ($dI_{\rm el}/dV$), is believed to be a possible tool in 
experiments to detect the photon-phonon-assisted multi-peaks in time-average 
transmission due to its proportionality to $T_{\rm tot(el)}(eV)$ at low temperature. 
So we illustrate in Fig.\,4 the calculated time-average current [(a)-(d)] and the 
differential conductance [(e)-(h)] for the case of bare level $\epsilon_{d}=1.0$ 
without ac field (a,e) and under ac irradiations with different frequencies 
$\Omega=0.5$ (b,f), $1.0$ (c,g), and $1.5$ (d,h) at a weak intensity $\Delta=0.4$. 
For simplicity, we assume the bias voltage is added symmetrically, i.e., $+V/2$ on 
the left lead and $-V/2$ on the right lead. Obviously, the time-average current 
shows steps indicating that new satellite-peak in transmission denoted by $(n,m)$ 
participate in conducting electrons. And correspondingly, there appears a peak in 
the $dI/dV$-$V$ curve located at the bias voltage 
$eV/2=\epsilon_{d}-\lambda^2/\omega_{ph}+n\hbar \omega_{ph}+m\hbar \Omega$. For the 
low frequency irradiation field $\Omega=0.5$, we observe a peak below the 
main-phonon peak $(0,0)$ resulted from the pure one-photon absorption process 
$(0,-1)$, which is nearly an elastic phonon process and has little inelastic 
contribution to the current. Moreover there is another peak between the main-phonon 
and the one-phonon peaks, which can be attributed to the combination contributions 
of a pure one-photon emission process $(0,1)$ and a one-photon-absorption-assisted 
one-phonon emission process $(1,-1)$. It possess of course inelastic contribution 
with a small portion.  	  

As pointed above, in order to observe more photon-phonon-assisted processes, a possible 
method is to increase the intensity of ac field. In Fig.\,5, we show the bias 
voltage-dependent time-average current and the differential conductance under several 
different microwave intensities $\Delta=0.4$, $0.8$, and $1.6$ for $\Omega=0.5$ (a,b) 
and $1.5$ (c,d). We find the following features in this figure: 1) Increasing 
irradiation intensity decreases the time-average current at higher bias voltage, but 
enhances the current at lower bias voltage; 2) The peak due to the pure phonon process 
$(0,0)$ is suppressed while the peak concerned with the photon-sbsorption processes 
$(0,-1)$ [Fig.\,5(b)] or $(1,-1)$ [(d)] grows up gradually with rising of ac intensity; 
3) These peaks due to photon-phonon combination processes have complicated relation with 
the ac intensity in the case of low-frequency irradiation field, which can be attributed 
to summation of the Bessel function with respect to the ac intensity in the time-average 
current formula Eq.\,(\ref{itot}).        

\begin{figure}[htb]
\includegraphics [width=8.cm,height=6.5cm,angle=0,clip=on] {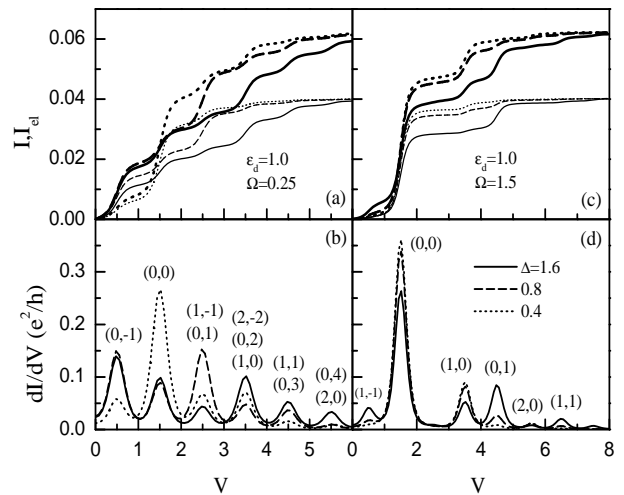}
\caption{The time-average current $I$ [(a,c)], and the corresponding differential 
conductance $dI/dV$ [(b,d)] vs bias voltage for the case of $\epsilon_{d}=1.0$ under 
several different irradiation intensities $\Delta=0.4$ (dotted line), $0.8$ (dashed 
line), and $1.6$ (solid line). (a,b) are ploted for $\Omega=0.5$, (c,d) for 
$\Omega=1.5$.} \label{fig5}
\end{figure}

Now we turn to study the variation of time-average current with the irradiation 
frequency at a given bias. Fig.\,6 depicts this MW spectroscopy for the QD with bare 
level $\epsilon_{d}=1.0$ at a low bias voltage $eV=0.2$. In this case, the low-frequency 
irradiation field induces a remarkable enhancement of the time-average current, in 
comparison with the dc current $I_{0}$ without irradiation, when electrons absorb energy 
of photon to overcome the energy gap between the left lead and the QD. The enhanced dc 
current is called pumped current.\cite{Wiel2} In Fig.\,6, we observe two obvious peaks 
with one order of magnitude higher than the dc current $I_{0}$, which are due to the 
one-photon- and two-photon-absorption induced pumping. But the positions of the two 
peaks marked by triangle symbols experience blue-shift 
$\Omega=(eV/2+\lambda^2/\omega_{ph}-\epsilon_{d})/m$ ($\Omega\simeq 0.71$ for $m=1$, and 
$\Omega\simeq 0.35$ for $m=2$) due to the polaron effect. More interestingly, we find 
two more evident pumped peaks $(1,-1)$ and $(2,-1)$ with smaller amplitude at 
high-frequency region, where the irradiation field can not only produce pumped current 
but also provide energy for electrons to emit phonons when electrons are pumped to 
tunnel through the molecule. These phonon-assisted pumped peaks locate at 
$\Omega=(eV/2+\lambda^2/\omega_{ph}-\epsilon_{d}-n\omega_{ph})/m$. And they have of 
course higher amplitudes in the presence of stronger irradiation. This addresses that 
the MW spectroscopy of the pumped current is another possible method to observe the 
photon-phonon-assisted resonant tunneling.

\begin{figure}[htb]
\includegraphics [width=6.cm,height=6cm,angle=0,clip=on] {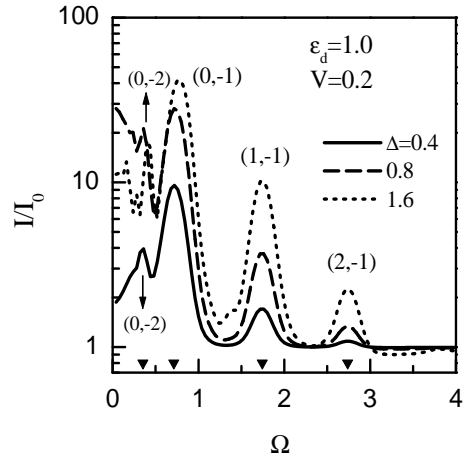}
\caption{The renormalized time-average total current $I$ vs the frequency of irradiation 
ac field for $\epsilon_{d}=1.0$ at the bias $eV=0.2$ under several different irradiation 
intensities $\Delta=0.4$ (solid line), $0.8$ (dashed line), and $1.6$ (dotted line). 
$I_{0}$ is the dc current of this system without irradiation.} \label{fig6}
\end{figure}

\section{Conclusions}

In this work we study the low-temperature time-dependent resonant tunneling through a 
single-molecular QD subjected to a dispersionless local phonon mode with a strong 
coupling to electrons. Due to the weak elastic parameters in molecular materials, the 
energy of boson field is matched with the MW irradiation field. So when electrons enter 
into the tunneling region, electrons can easily excite the phonon field with and without 
assistance of the ac field. Therefore, it is predicted that the combination influence of 
the phonon and photon fields will essentially change transport properties in the 
sigle-molecular QD.      

Owing to the strong EPI, the traditional perturbation theory is invalid for this problem 
even though it has been intensively applied for interpretation the classic transport 
phenomena in bulk semiconductor. Following the pioneer work of Bon\v ca and Trugman, we 
first transform this many-body problem exactly into a one-body scattering problem, by 
projecting the original Hamiltonian in the representation of electron-phonon coupled 
Fock space, i.e., the direct-product states of electron states and phonon number states. 
Based on this noninteracting Hamiltonian, we use the NGF approach to derive the 
time-dependent and the time-average transmission and current in the wide band 
approximation. For the sake of simplicity, we neglect the EPI-induced band narrowing in 
the present calculation. Moreover we adopt the adiabatic approximation for the 
irradiation MW field. 

Our numerical results show that time-average transmission vs incident electron energy 
$\omega$ displays additional peaks, besides original phonon absorption peaks, due to 
photon emission or absorption mediated processes as long as the irradiation frequency 
matches the condition $\omega=\epsilon_{d}-\lambda^2/\omega_{ph}+n\hbar 
\omega_{ph}+m\hbar \Omega$. We also point out that these new features will be observed 
in experiments by measuring the nonlinear differential conductance and the MW 
spectroscopy of pumped current at proper irradiation intensity.

\begin{acknowledgments} 

B. Dong and H. L. Cui are supported by the DURINT Program administered by the US Army 
Research Office. X. L. Lei is supported by Major Projects of National Natural Science 
Foundation of China (10390162 and 90103027), the Special Founds for Major State Basic 
Research Project (G20000683) and the Shanghai Municipal Commission of Science and 
Technology (03DJ14003).

\end{acknowledgments}


\begin{thebibliography}{99}

\bibitem{Chen}{J. Chen, M. Reed, A. Rawlett, and J. Tour, Science {\bf 286}, 1550 
(1999).}

\bibitem{Park}{H. Park, J. Park, A. Lim, E. Anderson, A. Allvisatos, and P. McEuen, 
Nature {\bf 407}, 57 (2000).}

\bibitem{Ventra}{M. Ventra, S. G. Kim, S. Pantelides, and N. Lang, Phys. Rev. Lett. {\bf 
86}, 288 (2001).}

\bibitem{Zhitenev}{N. B. Zhitenev, H. Meng, and Z. Bao, Phys. Rev. Lett. {\bf 88}, 
226801 (2002); L.H. Yu and D. Natelson, Nano Letters, {\bf 4}, 79 (2004).}

\bibitem{Fujisawa}{T. Fujisawa, T. H. Oosterkamp, W. G. van der Wiel, B. W. Broer, R. 
Aguado, S. Tarucha, Leo P. Kouwenhoven, Science {\bf 282}, 932 (1998).}

\bibitem{Turley}{P. J. Turley and S. W. Teitsworth, Phys. Rev. B, {\bf 44}, 3199 
(1991).}

\bibitem{Brandes}{T. Brandes and B. Kramer, Phys. Rev. Lett. {\bf 83}, 3021 (1999); T. 
Brandes, N. Lambert, Phys. Rev. B {\bf 67}, 125323 (2003)}

\bibitem{Bose}{D. Bose and H. Schoeller, Europhys. Lett. {\bf 54}, 668 (2001); K. D. 
McCarthy, N. Prokof'ev, and M. T. Tuominen, cond-mat/0205419; A. Mitra, I. Aleiner, and 
A. J. Millis, cond-mat/0302132, (2003).}

\bibitem{Wingreen}{N. S. Wingreen, K. W. Jacobsen, and J. W. Wilkins, Phys. Rev. Lett. 
{\bf 61}, 1396 (1988); Phys. Rev. B {\bf 40}, 11834 (1989).}

\bibitem{Lundin}{U. Lundin and R. McKenzie, Phys. Rev. B {\bf 66}, 75303 (2002); J. X. 
Zhu and A. V. Balatsky, Phys. Rev. B {\bf 67}, 165326 (2003).}

\bibitem{Flensberg}{K. Flensberg, Phys. Rev. B {\bf 68}, 205323 (2003); S. Braig and K. 
Flensberg, Phys. Rev. B {\bf 68}, 205324 (2003).}

\bibitem{Wiel}{W. G. van der Wiel, S. De Franceschi and J. M. Elzerman, T. Fujisawa, S. 
Tarucha, L. P. Kouwenhoven, Rev. Modern Phys. {\bf 75}, 1 (2003).}

\bibitem{Jauho}{A. P. Jauho, N. S. Wingreen, and Y. Meir, Phys. Rev. B {\bf 50}, 5528 
(1994).}

\bibitem{Bonca}{J. Bon\v ca and S. A. Trugman, Phys. Rev. Lett. {\bf 75}, 2566 (1995).}

\bibitem{Holstein}{T. Holstein, Ann. Phys. (N.Y.S.) {\bf 8}, 325 (1959).}

\bibitem{SSH}{W. P. Su, J. R. Schrieffer, and A. J. Heeger, Phys. Rev. Lett. {\bf 42}, 
1698 (1979).}

\bibitem{Bonca2}{J. Bon\v ca and S. A. Trugman, Phys. Rev. Lett. {\bf 79}, 4874 (1997); 
K. Haule and J. Bonc\"a, Phys. Rev. B {\bf 59}, 13087 (1999).}

\bibitem{Ness}{H. Ness and A. J. Fisher, Phys. Rev. Lett. {\bf 83}, 452 (1999); E. G. 
Emberly and G. Kirczenow, Phys. Rev. B {\bf 61}, 5740 (2000).}

\bibitem{Langreth}{D. C. Langreth, in {\it Linear and Nonlinear Electron Transport in 
Solids, Nato ASI, Series B} vol. 17, Ed. J. T. Devreese and V. E. Van Doren (Plenum, New 
York, 1976).}

\bibitem{Ng}{T. K. Ng, Phys. Rev. Lett. {\bf 76}, 487 (1996).}

\bibitem{Lee}{P. A. Lee and D. S. Fisher, Phys. Rev. Lett. {\bf 47}, 882 (1981).}

\bibitem{Wiel2}{For review, see W. G. van der Wiel, T. H. Oosterkamp, S. De Franceschi, 
C. J. P. M. Harmans, and L. P. Kouwenhoven, in {\it Strongly Correlated Fermions and 
Bosons in Low-Dimensional Disordered Systems}, Ed. I. V. Lerner, B. L. Altshuler, and V. 
I. Fal'ko (Kluwer Academic, Dordrecht, 2002), pp. 43-68, and references therein.}

\end{thebibliography}
\end{document}